\documentstyle[12pt,fleqn]{article}
\newdimen\mathindent\mathindent=6mm
\newcommand{\fl}{\hspace*{-\mathindent}}%
\newcommand{\tqs}{\hspace*{25pt}}%
\newcommand{\case}[2]{{\textstyle\frac{#1}{#2}}}
\newif\ifnumbysec
\newcommand{\eref}[1]{(\ref{#1})}%
\def\eqalign#1{\null\vcenter{\def\\{\cr}\openup\jot
  \ialign{\strut$\displaystyle{##}$\hfil&$\displaystyle{{}##}$\hfil
      \crcr#1\crcr}}\,}
\def\cases#1{%
     \left\{\,\vcenter{\def\\{\cr}\normalbaselines\openup1\jot
     \ialign{\strut$\displaystyle{##}\hfil$&\tqs
     \rm##\hfil\crcr#1\crcr}}\right.}%

\def\ack{%
{\bigskip\noindent\normalsize\bf~{Acknowledgments}\raggedright\par\medskip}}%

\begin{document}

\title{An application of the $3$-dimensional $q$-deformed harmonic oscillator
to the nuclear shell model}
\author{
P P Raychev\dag\S\thanks{\tt email:~ Raychev@bgcict.acad.bg
                                   ~~Raychev@inrne.acad.bg},
R P Roussev\S,\thanks{\tt email:~ Rousev@inrne.acad.bg}
N Lo Iudice\dag\thanks{\tt email:~ Nicola.Loiudice@na.infn.it}
~and P A Terziev\S  }
\date{1 July 1997}
\maketitle

{\noindent\small
\hspace*{10mm}
\dag\ Dipartimento di Scienze Fisiche, Universit\`a di Napoli ``Federico II''\\
\hspace*{10mm}
Mostra d' Oltremare, Pad. 19, I-80125 Napoli, Italy\\
\hspace*{10mm}
\S\ Institute for Nuclear Research and Nuclear Energy\\
\hspace*{10mm}
Bulgarian Academy of Sciences, 72 Tzarigrad Road, BG-1784 Sofia, Bulgaria}

\begin{abstract}
An analysis of the construction of a $q$-deformed version of the 3-dimensional
harmonic oscillator, which is based on the application of $q$-deformed
algebras, is presented. The results together with their applicability to the
shell model are compared with the predictions of the modified harmonic
oscillator.
\end{abstract}

\section{\normalsize\bf
Introduction}

	The assumption of an independent particle motion is the basis of
	shell model in finite nuclei. The average field is correctly
	described by the Woods-Saxon potential $V_{WS}({\bf r})$.
	Since, however, the corresponding eigenvalue equation can be
	solved only numerically, it is desirable for many purposes
	to make use of the so called modified harmonic oscillator (MHO)
	Hamiltonian, which reproduces approximately
	the energy spectrum of the WS potential and at the same time
	has a simple analytical solution.
	The MHO was suggested by Nilsson \cite{Ni55} and studied in
	detail in \cite{Gus67,BR85,NR95}.  It has the form
\begin{equation}\label{eq:i01}
V_{MHO} = \frac{1}{2}\hbar \omega \rho^2 -
\hbar\omega k \left\{\mu\left({\bf L}^2 -
\left\langle{\bf L}^2\right\rangle_N\right) + 2{\bf L.S}\right\},
\quad \rho = \sqrt{\frac{M \omega}{\hbar}}\, r
\end{equation}
	where ${\bf L}^2$ is the square of the angular momentum and
	${\bf L.S}$ is the spin-orbit interaction.

	As it is well known, the subtraction of the average value of
	$\langle{\bf L^2}\rangle$, taken over each $N$-shell
\begin{equation}\label{eq:i02}
\left\langle {\bf L}^2 \right\rangle_N = \frac{N(N+3)}{2}
\end{equation}
	avoids the shell compression induced by the ${\bf L}^2$ term,
	leaving the ``center of gravity'' of each $N$-shell unchanged.
	The $V_{MHO}$ (without the spin-orbit term ${\bf L.S}$) reproduces
	effectively the Woods - Saxon radial potential.
	A better agreement with the experimental data can be achieved if the
	parameters of the potential are let to vary smoothly from shell to
	shell.

	It should be noted, however, that from mathematical point of view
	the introduction of ${\bf L^2}$-dependent term (and the corresponding
	correction $\langle{\bf L}^2\rangle_N$) in the modified harmonic
	oscillator potential is not so innocent, as it look at first sight,
	because this term depends on the state in which the particle occurs,
	and it this sense the potential \eref{eq:i01} is ``non-local'' and
	``deformable'' (with variable deformation).  This effect is amplified
	by the fact, that the constant $\mu^{\prime}=k\mu$ varies from shell
	to shell.

	The inclusion of ${\bf L}^2$-dependent term together with the
	corresponding correction $\langle{\bf L}^2\rangle_N$ and of the
	$N$-dependence of $\mu^{\prime}$ can be considered as a first step
	for the introduction of some non-trivial `` deformations'' into the
	Hamiltonian of the 3-dimensional harmonic oscillator.  There is,
	however, an alternative approach to the description of the space
	extended and deformable many body systems based on the application of
	the $q$-deformed algebras.

	In nuclear physics, the $su_q(2)$ quantum algebra has been used
	with promising results for the description of rotational bands in the
	deformed and superdeformed nuclei \cite{Ray2} - \cite{Kib}.
	Symmetries based on quantum algebras have been adopted
	successfully for studying rotational \cite{BRRS,ETV}, vibrational
	\cite{BRF, BAR} and rotational - vibrational spectra
	\cite{BD}-\cite{CGY} of diatomic molecules. Quantum groups
	extensions of some simple nuclear models \cite{DLMV}-\cite{BDFRR}
	have also been shown to lead to interesting results.

	The aim of this paper is to construct a Hamiltonian of the
	$q$-deformed 3-dimensional harmonic oscillator in terms of the
	$q$-deformed boson operators \cite{Bie,Mac}. It will be shown that,
	in the $q \to 1$ limit, its eigenvalues ${}^qE$ coincide with the
	eigenvalues of the modified harmonic oscillator (MHO) without a
	spin-orbit term. More specifically, we  represent $q$ in the
	form $q=e^{\tau}$ and  show that the expansion of ${}^qE$ in series
	of $\tau$ yields a Hamiltonian having the same structure as the one
	given by (\ref{eq:i01}), with $\tau$ playing the role of
	$\mu^{\prime}= k\mu$.  It may be worth to stress at this point,
	that the construction of the
	Hamiltonian ${}^qH_{3 dim}$ is a non-trivial problem since, from
	a physical point of view, it is necessary to separate explicitly the
	$q$-deformed angular momentum term $^q{\bf L}^2$.

	In Section 2 we introduce $q$-deformed $so_q(3)$ generators
	($q$-analogs of the angular momentum operators), which act in the
	space of the most symmetric representation of the algebra $u_q(3)$,
	which, as a matter of fact, is the space of the $q$-deformed
	3-dimensional harmonic oscillator \cite{J1,J2,J3}.

	In Section 3 we suggest an expression for the Hamiltonian of the
	$q$-deformed 3-dimensional harmonic oscillator, which is invariant
	with respect to the $q$-deformed angular momentum algebra. The
	vibrational and rotational degrees of freedom of this Hamiltonian are
	clearly separated, which is in agreement with the classical situation
	and in the limit $q\to 1$ it coincides with the classical
	Hamiltonian (without the correction term \eref{eq:i02}). In the case
	of small deformations of the algebra its spectrum reproduces very
	well the spectrum of the 3-dimensional MHO with the correction
	(\ref{eq:i02}).

	The next step (Section 4) is to introduce the $q$-deformed spin-orbit
	term ${\cal L}^{(q)}\cdot{\cal S}^{(1/q)}$ and to show that in the
	case of a small deformations this term introduces small, but
	essential corrections to the standard spin-orbit term
	${\bf L}\cdot{\bf S}$.

\section{\normalsize\bf
$so_q(3)$-algebra and $so_q(3)$-basis states} \label{sec:so3}

	We shall use the three independent $q$-deformed boson
	operators $b_i$ and $b_i^\dagger$ ($i=+,0,-$),
	which satisfy the commutation relations \cite{Bie,Mac}
\begin{equation}\label{eq:s01}
[N_i,b_i^\dagger ] = b_i^\dagger\qquad
[N_i,b_i] = -b_i\qquad
b_i b_i^\dagger -  q^{\pm1} b_i^\dagger b_i = q^{\mp N_i}
\end{equation}
	where $N_i$ are the corresponding number operators.

	The elements of the $so_q(3)$
	algebra (i.e. the angular momentum operators) acting in the Fock
	space of the totally symmetric representations $[N,0,0]$ of $u_q(3)$
	have been derived by Van der Jeugt \cite{J1,J2,J3}.
	A simplified form of the subalgebra $so_q(3)\subset u_q(3)$,
	can be obtained by introducing the operators \cite{RRTL}
\begin{eqnarray}\label{eq:s02}
B_0 = q^{-\frac{1}{2}N_0}b_0 \qquad
&& B_0^\dagger = b_0^\dagger q^{-\frac{1}{2}N_0}\nonumber\\
B_i = q^{N_i+\frac{1}{2}} b_i \sqrt{\frac{[2 N_i]}{[N_i]}} \qquad
&& B_i^\dagger = \sqrt{\frac{[2 N_i]}{[N_i]}} b_i^\dagger q^{N_i+\frac{1}{2}}
\qquad i=+,-
\end{eqnarray}
	where the $q$-numbers $[x]$ are defined as
\begin{displaymath}
[x] = (q^x-q^{-x})/(q-q^{-1})
\end{displaymath}
	and we shall consider only real values for the deformation parameter
	$q$. These new operators satisfy the usual commutation relations
\begin{equation}\label{eq:s03}
[N_i,B^{\dagger}_i] = B_i^{\dagger}\qquad
[N_i,B_i] = -B_i .
\end{equation}
	In the Fock space, spanned on the normalized
	eigenvectors of the excitation number operators $N_{+}, N_{0}, N_{-}$,
	they satisfy the relations
\begin{eqnarray}\label{eq:s04}
B^{\dagger}_{0} B_{0} = q^{- N_{0}+1}[N_{0}]\qquad
&& B_{0} B^{\dagger}_{0} = q^{-N_{0}}[N_{0}+1]\nonumber\\
B_i^\dagger B_i = q^{2N_i-1}  [2N_i] \qquad
&& B_i B_i^\dagger =
q^{2N_i+1} [2N_i+2]\qquad i=+,-
\end{eqnarray}
	and hence, the commutation relations
\begin{equation}\label{eq:s05}
[B_{0},B^{\dagger}_{0}]= q^{-2N_{0}}\qquad
[B_i,B_i^\dagger] = [2] q^{4 N_i+1}\qquad i=+,-
\end{equation}
	It was shown in \cite{RRTL} that the angular momentum
	operators defined in \cite{J1,J2}, when are expressed
	in terms of the modified operators \eref{eq:s02},
	take the simplified form
\begin{eqnarray}\label{eq:s06}
L_{0} &=& N_{+}-N_{-}\nonumber\\
L_{+} &=& q^{-L_{0}+{1\over 2}}B^{\dagger}_{+} B_{0} + q^{L_{0}-{1\over 2}}
B^{\dagger}_{0} B_{-}\nonumber\\
L_{-} &=& q^{-L_{0}-{1\over 2}}B^{\dagger}_{0} B_{+} + q^{L_{0}+{1\over 2}}
B^{\dagger}_{-} B_{0}.
\end{eqnarray}
	and satisfy the standard $so_q(3)$ commutation relations
\begin{equation}\label{eq:s07}
[L_0,L_{\pm}] = \pm L_{\pm} \qquad [L_+,L_-] = [2L_0]
\end{equation}
	As it has been discussed in detail in \cite{J1} the algebra
	\eref{eq:s06} is not a subalgebra of $u_q(3)$ considered as
	Hopf algebras, but as $q$-deformed enveloping algebras. However,
	in our context, working in the Fock space (constructed from three
	independent $q$-bosons), we restrict ourselves
	to the symmetric representation of $u_q(3)$, where the embedding
	$so_q(3)\subset u_q(3)$ is valid.

	The Casimir operator of $so_q(3)$ can be written in the form \cite{Sm}
\begin{eqnarray}\label{eq:s08}
C_2^{(q)} &=& \frac{1}{2}\left\{L_{+}L_{-} + L_{-}L_{+} + [2][L_0]^2\right\}
\nonumber\\
&=& L_{-}L_{+} + [L_0][L_0 + 1] = L_{+}L_{-} + [L_0][L_0 - 1] .
\end{eqnarray}
	One can also define $q$-deformed
	states $|n \ell m\rangle_q$ satisfying the eigenvalue
	equations
\begin{eqnarray}\label{eq:sh01}
{}^q{\bf L}^2 |n \ell m \rangle_q &=& [\ell][\ell+1] |n \ell m\rangle_q
\nonumber\\
L_0 |n \ell m \rangle_q &=& m |n \ell m\rangle_q \nonumber\\
N |n \ell m \rangle_q &=& n |n \ell m\rangle_q
\end{eqnarray}
	Here $^q{\bf L}^2 \stackrel{def}{=} C^{(q)}_2$ is the ``square'' of the
	$q$-deformed angular momentum and $ N = N_+ + N_0 + N_-$ is the
	total number operator for the $q$-deformed bosons. These
	states have the form
\begin{eqnarray}\label{eq:b16}
\fl && |n \ell m\rangle_q = q^{\frac{1}{4}(n-\ell)(n+\ell+1)-\frac{1}{2}m^2}
\sqrt{\frac{[n-\ell]!![\ell+m]![\ell-m]![2\ell+1]}{[n+\ell+1]!!}} \times
\nonumber\\
\fl && \qquad\times
\sum_{t=0}^{(n-\ell)/2} \sum_{p=\max(0,m)}^{\lfloor(\ell+m)/2\rfloor}
\frac{(-1)^t q^{-(n+\ell+1)t}}{[2t]!![n-\ell-2t]!!}
\frac{(B^\dagger_{+})^{p+t}}{[2p]!!}
\frac{(B^\dagger_{0})^{n+m-2p-2t}}{[n+m-2p]!}
\frac{(B^\dagger_{-})^{p+t-m}}{[2p-2m]!!}
|0\rangle
\end{eqnarray}
	where $|0\rangle$ is the vacum state, $[n]!=[n][n-1]\ldots [1]$,~
	$[n]!!=[n][n-2]\ldots [2]\;\mbox{or}\;[1]$,
	and, as shown in \cite{J1,J2}, form a basis for
	the most symmetric representation $[n,0,0]$ of $u_q(3)$,
	corrresponding to the $u_q(3) \supset so_q(3)$ chain.

\section{\normalsize\bf
$so_q(3)$ vector operators and the Hamiltonian of the
3-dimensional $q$-de\-for\-med harmonic oscillator}

	Here we shall recall some well known definitions about the
	$q$-deformed tensor operators within the framework of the algebra
	$so_q(3)$. An irreducible tensor operator of rank $j$
	with parameter $q$ according to the algebra $so_q(3)$ is a set of
	$2j+1$ operators ${\cal T}^{(q)}_{j m}$, satisfying the relations
\begin{eqnarray} \label{eq:tensor}
&& [L_0,{\cal T}^{(q)}_{j m}] = m\,{\cal T}^{(q)}_{j m} \nonumber\\{}
&& [L_{\pm},{\cal T}^{(q)}_{j m}]_{q^m} q^{L_0} =
\sqrt{[j \mp m][j \pm m + 1]}\,{\cal T}^{(q)}_{j m\pm1}
\end{eqnarray}
	where, in order to express the adjoint action of the generators
	$L_0,L_{\pm}$ of $so_q(3)$ on the components of the tensor operator
	${\cal T}^{(q)}_{j m}$, we  use the usual notation of the
	$q$-commutator
\begin{equation}\label{eq:qcom}
[A,B]_{q^\alpha} = A B - q^{\alpha} B A  .
\end{equation}
	By $\widetilde{\cal T}^{(q)}_{j m}$ we denote
	the conjugate irreducible $q$-tensor operator
\begin{equation} \label{eq:ttensor}
\widetilde{\cal T}^{(q)}_{j m} = (-1)^{j-m}q^{-m}{\cal T}^{(q)}_{j, -m} \ ,
\end{equation}
	which satisfies the relations
\begin{eqnarray} \label{eq:rctensor}
&& [\widetilde{\cal T}^{(q)}_{j m},L_{0}] = m\;\widetilde{\cal T}^{(q)}_{j m}
\nonumber\\
&& q^{L_0}[\widetilde{\cal T}^{(q)}_{j m},L_{\pm}]_{q^m} =
\sqrt{[j\pm m][j\mp m+1]}\;\widetilde{\cal T}^{(q)}_{j, m\mp 1} \ .
\end{eqnarray}
	Then the operator
\begin{equation} \label{eq:conjtensor}
{\cal P}^{(q)}_{j m} = (\widetilde{\cal T}^{(q)}_{j m})^\dagger =
(-1)^{j-m} q^{-m} ({\cal T}^{(q)}_{j, -m})^\dagger
\end{equation}
	where ${}^\dagger$ denotes hermitian conjugation, transforms
	in the same way \eref{eq:tensor} as the tensor ${\cal T}^{(q)}_{j m}$,
	i.e.  ${\cal P}^{(q)}_{j m}$ also is an irreducible $so_q(3)$
	tensor operator of rank $j$.

	Let ${\cal A}^{(q_1)}_{j_1 m_1}$ and ${\cal B}^{(q_2)}_{j_2 m_2}$
	be two irreducible tensor operators.
	We shall define the tensor and
	scalar product of these tensor operators following some of the
	prescriptions, summarized, for example, in \cite{Sm}.  One can
	introduce the following operator
\begin{equation} \label{eq:prod}
\left[{\cal A}^{(q_1)}_{j_1}\times{\cal B}^{(q_2)}_{j_2}\right]^{(q_3)}_{j m}
= \sum_{m_1,m_2} {}^{q_3} C_{j_1 m_1, j_2 m_2}^{j m}
{\cal A}^{(q_1)}_{j_1 m_1} {\cal B}^{(q_2)}_{j_2 m_2}
\end{equation}
	where ${}^{q_3} C_{j_1 m_1, j_2 m_2}^{j m}$ are the Clebsch-Gordan
	coefficients corresponding to the deformation parameter $q_3$. In
	general the deformation parameters $q_1, q_2$ and $ q_3$ can be
	arbitrary. It turns out, however, if one imposes the condition
	that the left hand side of (\ref{eq:prod}) transforms as an
	irreducible $q$-tensor of rank $j$
	in a way, that the Wigner-Eckhart theorem can be
	applied to (\ref{eq:prod}) as a whole, not all of the combinations of
	$q_1, q_2$ and $q_3$ are allowed.

	If the tensors ${\cal A}^{(q_1)}_{j_1}$ and ${\cal B}^{(q_2)}_{j_2}$
	depend on one and same variable and act on a single vector,
	which depend on the same variable, the mentioned requirement will
	be satisfied only if $q_1 = q_2 = q$ and $q_3=1/q$, i.e. the operator
\begin{equation} \label{eq:prod1}
\left[{\cal A}^{(q)}_{j_1}\times{\cal B}^{(q}_{j_2} \right]^{(1/q)}_{j m}
= \sum_{m_1,m_2} {}^{1/q}\!C_{j_1 m_1, j_2 m_2}^{j m}
{\cal A}^{(q)}_{j_1 m_1} {\cal B}^{(q)}_{j_2 m_2}
\end{equation}
	transforms as an irreducible $q$-tensor of rank $j$ according to the
	algebra $so_q(3)$ . Then, the definition (\ref{eq:prod1}) is
	in agreement with the property
\begin{eqnarray} \label{single}
\fl \langle\alpha',\ell'\| [{\cal A}^{(q)}_{j_1} \times
{\cal B}^{(q)}_{j_2} ]^{(1/q)}_{j} \|\alpha,\ell\rangle
&=& \sqrt{[2j+1]}\sum_{\alpha'',\ell''} (-1)^{\ell+j+\ell'}
\left\{\begin{array}{ccc}\ell & j & \ell' \\ j_1 & \ell'' & j_2 \end{array}
\right\}_{q} \times\nonumber\\
\fl &&\times
\langle\alpha',\ell'\|{\cal A}^{(q)}_{j_1}
\|\alpha'',\ell''\rangle
\langle\alpha'',\ell''\|{\cal B}^{(q)}_{j_2}
\|\alpha,\ell\rangle
\end{eqnarray}
	which is a $q$-analogue of the well known classical identity.

	In Section 4 we shall consider also the scalar product of
	two tensor operators depending on two different variables $(1)$ and
	$(2)$ and acting on different vectors, depending on these different
	variables.  In this particular case the {\it scalar product} of
	irreducible $q$ and $q^{-1}$ -- tensor operators
	${\cal A}^{(q)}_{j}(1),~{\cal B}^{(1/q)}_{j}(2)$ with the same rank,
	acting on {\it different vectors} $(1)$ and $(2)$,
	will be given by means of the following definition
\begin{eqnarray} \label{eq:scalar}
({\cal A}^{(q)}_{j}(1)\cdot{\cal B}^{(1/q)}_{j}(2))^q
&=& (-1)^j\sqrt{[2j+1]}
\left[{\cal A}^{(q)}_{j}(1)\times{\cal B}^{(1/q)}_{j}(2)\right]^{(q)}_{0 0}
\nonumber\\
&=& \sum_{m} (-q)^m {\cal A}^{(q)}_{j m}(1)~ {\cal B}^{(1/q)}_{j -m}(2) \ .
\end{eqnarray}

	It should be noted that the angular momentum operators (\ref{eq:s06})
	satisfy the commutation relations (\ref{eq:s07}), but {\it are not}
	tensor operators in the sense of definition (\ref{eq:tensor}).
	One can form, however, the angular momentum operators
\begin{eqnarray} \label{eq:Ltensor} \nonumber
{\cal L}^{(q)}_{\pm1} &=& \mp\frac{1}{\sqrt{[2]}} q^{-L_0} L_{\pm} \\
{\cal L}^{(q)}_{0} &=& \frac{1}{[2]}\left\{q[2L_0] +
(q-q^{-1})L_{-}L_{+}\right\} = \nonumber\\
&=& \frac{1}{[2]}\left\{q[2L_0] +
(q-q^{-1})\left( C_2^{(q)} - [L_0][L_0+1] \right) \right\}
\end{eqnarray}
	which are tensors of 1-st rank, i.e $so_q(3)$ vectors\footnote{
	The situation in the ``standard'' theory of angular momentum is
	analogous. Indeed, from  the operators $L_0 = L_3, \,
	L_{\pm}=L_1~\pm~iL_2$, which are
	not spherical tensors, one can construct the operators
	$J_{\pm} = \mp \textstyle\frac{1}{\sqrt{2}}(L_1 \pm i L_2), J_0=L_0$,
	which are the components of a spherical tensor of 1-st rank according
	the standard $so(3)$-algebra.}
	(see for example \cite{Sm}).

	One can easily check that the operators $B^{\dagger}_0,
	B^{\dagger}_{\pm}$ and $B_0, B_{\pm}$ do not form a spherical vector.
	As shown in \cite{RRTL}, one can define
	a vector operator $T^\dagger_m$ of the form
\begin{eqnarray}\label{eq:vector1}
T^{\dagger}_{+1} &=& {1\over \sqrt{[2]}} B^{\dagger}_{+}
q^{-2N_{+} + N - {1\over 2}}\nonumber\\
T^{\dagger}_{0} &=& B^{\dagger}_{0}  q^{-2N_{+} + N}\nonumber\\
T^{\dagger}_{-1} &=& {1\over \sqrt{[2]}}\left\{ B^{\dagger}_{-}
q^{2 N_{+} - N - {1\over 2}} - (q - q^{-1}) B_{+} {(B^{\dagger}_{0})}^2
q^{-2 N_{+} + N + {3\over 2}}\right\}.
\end{eqnarray}
	The corresponding expressions for the conjugate operators
	$\widetilde{T}_m = (-1)^m q^{-m} (T^\dagger_{-m})^\dagger$ are
\begin{eqnarray}\label{eq:vector2}
\widetilde{T}_{+1} &=& -{1\over \sqrt{[2]}}\left\{
q^{2 N_{+} - N - {3\over 2}} B_{-} - (q-q^{-1}) q^{-2N_{+} + N+ {1\over 2}}
B^{\dagger}_{+} {(B_{0})}^2 \right\}\nonumber\\
\widetilde{T}_{0} &=& q^{-2N_{+} + N} B_{0}\nonumber\\
\widetilde{T}_{-1} &=& -{1\over \sqrt{[2]}}q^{-2N_{+}+N+{1\over 2}} B_{+} .
\end{eqnarray}
	One can easily check, that vector operators ${\bf T}^{\dagger}$ and
	$\widetilde{\bf T}$ satisfy the commutation relation
\begin{equation}\label{eq:to1}
\eqalign{
[\widetilde{T}_{-1},T^\dagger_{+1}]_{q^{-2}} = -q^{2N+1} \\
[\widetilde{T}_0,T^\dagger_0] = q^{2N} + q^{-1}(q^2-q^{-2})
T^\dagger_{+1}\widetilde{T}_{-1} \\
[\widetilde{T}_{+1},T^\dagger_{-1}]_{q^{-2}} = -q^{2N-1} + q^{-1}(q^2-q^{-2})
\!\left\{T^\dagger_0\widetilde{T}_0 +
(q-q^{-1}) T^\dagger_{+1}\widetilde{T}_{-1}
\right\} \\ }
\end{equation}
	and
\begin{eqnarray}\label{eq:to2}
\eqalign{
[\widetilde{T}_0,T^\dagger_{+1}] = 0 \\
[\widetilde{T}_{+1},T^\dagger_{+1}]_{q^2} = 0 \\
[\widetilde{T}_{+1},T^\dagger_0] = (q^2-q^{-2})
T^\dagger_{+1}\widetilde{T}_0 \\ }
&\qquad&
\eqalign{
[\widetilde{T}_{-1},T^\dagger_0] = 0 \\
[\widetilde{T}_{-1},T^\dagger_{-1}]_{q^2} = 0 \\
[\widetilde{T}_0,T^\dagger_{-1}] = (q^2-q^{-2})
T^\dagger_0\widetilde{T}_{-1} \\ }
\end{eqnarray}
Unlike operators (\ref{eq:s02}), the commutators
$[\widetilde T_m, \widetilde T_n]$ and $ [T^{\dagger}_m,T^{\dagger}_n]$
do not vanish, but are equal to
\begin{eqnarray} \label{eq:to3}
\eqalign{
[T^\dagger_{+1},T^\dagger_0]_{q^2} = 0 \\
[T^\dagger_0,T^\dagger_{-1}]_{q^2} = 0 \\
[T^\dagger_{+1},T^\dagger_{-1}] = (q-q^{-1}) (T^\dagger_0)^2 \\ }
&\qquad&
\eqalign{
[\widetilde{T}_0,\widetilde{T}_{-1}]_{q^2} = 0 \\
[\widetilde{T}_{+1},\widetilde{T}_0]_{q^2} = 0 \\
[\widetilde{T}_{+1},\widetilde{T}_{-1}] = (q-q^{-1}) (\widetilde{T}_0)^2 \\ }
\end{eqnarray}
	which is in agreement with the results obtained in \cite{Q2}.

	It should be noted that the angular momentum operator ${\cal
	L}^{(q)}_M$ (considered as a vector according $so_q(3)$) can be
	represented in the form
\begin{equation}\label{eq:lo13}
{\cal L}^{(q)}_M = - \sqrt{\frac{[4]}{[2]}} \left[ T^{\dagger} \times
\widetilde{T} \right]_{1M}^{(1/q)} = - \sqrt{\frac{[4]}{[2]}} \sum_{m,n}
{}^{1/q} C_{1m,1n}^{1M} \,T^{\dagger}_m \widetilde T_n
\end{equation}
	Its ``square'' differs from $C^{(q)}_2$ and equals to \cite{Sm}
\begin{eqnarray}\label{eq:lo2}\nonumber
({\cal L}^{(q)})^2 \equiv {\cal L}^{(q)}\cdot {\cal L}^{(q)} &=&
\sum (-q)^{-M} {\cal L}^{(q)}_M{\cal L}^{(q)}_{-M} \\
&=&\frac{2}{[2]}C^{(q)}_2 + \left(\frac{q - q^{-1}}{[2]}\right)^2
(C^{(q)}_2)^2
\end{eqnarray}
	This difference, however, is not essential. Let us consider, for
	instance, the  expectation values of the scalar operator
	(\ref{eq:lo2}).  It has the form
\begin{equation}
{}_q\langle\ell m|{\cal L}^{(q)}\cdot{\cal L}^{(q)}|\ell m\rangle_q =
\frac{[2\ell][2\ell+2]}{[2]^2}=[\ell]_{q^2} [\ell+1]_{q^2}
\end{equation}
	In this sense the replacement of ${}^q {\bf L}^2 \equiv C^{(q)}_2$ with
	${\cal L}^{(q)}\cdot {\cal L}^{(q)}$ is equivalent to the replacement
	$q \to q^2$ and leads to the renormalization of some constant.  For
	this reason we shall accept as the square of the {\it physical}
	angular momentum the quantity $^q{\bf L}^2\equiv C^{(q)}_2$, whose
	eigenvalues are $[\ell][\ell+1]$.

	Now let us consider the scalar operator, constructed in terms of
	$T^{\dagger}_M$ and $\widetilde T_M$. We have
\begin{eqnarray}\label{eq:ha01} \nonumber
X_{0}^{(q)}
&=& -\sqrt{[3]}[T^\dagger\times \widetilde{T}]^{(1/q)}_{00}  \\
&=& -q^{-1}T^\dagger_{+1}\widetilde{T}_{-1}
+ T^\dagger_{0}\widetilde{T}_{0}
- q T^\dagger_{-1}\widetilde{T}_{1}
\end{eqnarray}
	We define the Hamiltonian of the three dimensional
	$q$-deformed oscillator as
\begin{equation}\label{eq:ha01a}
{}^qH_{3dim} = \hbar \omega_0 X_0 = \hbar \omega_0 \left(
-q^{-1}T^\dagger_{+1}\widetilde{T}_{-1} +
T^\dagger_{0}\widetilde{T}_{0} - q T^\dagger_{-1}\widetilde{T}_{1} \right)
\end{equation}
	The motivations for such an ansatz are:

	1. The operator so defined is an $so_q(3)$ scalar, i.e. it is
	simultaneously
	measurable with the physical $q$-deformed angular momentum square
	$^q {\bf L}^2$
	and the z-projection $L_0$;

	2.  Only this $so_q(3)$-scalar has the property of being the sum
	of terms, each containing
	an equal number of creation and annihilation operators (i.e.
	conserves the number of bosons);

	3. In the limit $q \to 1$ \eref{eq:ha01a} goes into
\[ \lim_{q\to 1} {}^q H_{3dim} =
\hbar \omega_0 \left(a^{\dagger}_{+1} a_{+1} +
a^{\dagger}_{0} a_{0} + a^{\dagger}_{-1} a_{-1} \right) \]
	where $[a_{m},a^{\dagger}_{n}] =  \delta_{m n}$, i.e. in this limit
	(\ref{eq:ha01a}) coincides with the Hamiltonian of the 3-dimensional
	(spherically symmetric) harmonic oscillator up to an additive constant.
	It should be underlined that the Hamitonian (\ref{eq:ha01a}) does not
	commute with the square of the ``classical'' (or ``standard'') angular
	momentum ${\bf L}^2$. This is due to the fact that the $q$-deformed
	oscillator is really space deformed and the ``standard'' quantum
	number $l$ is not a good quantum number for this system. On the other
	hand ${}^q {\bf L}^2$ commutes with ${}^qH_{3 dim}$ and the
	``quantum angular momentum'' $\ell$ can be used for the
	classification of the states of the $q$-oscillator. However the
	projection of the standard angular momentum on the $z$ axis $l_z$
	coincides with the quantum projectoin $\ell_z$, i.e. it is a good
	quantum number.

	Equation (\ref{eq:ha01a}) can be cast in a simpler and
	physically more transparent form. Indeed, making use of
	the third component of the $q$-deformed angular
	momentum (considered as $so_q(3)$ vector)
\begin{eqnarray}\label{eq:ha02}
{\cal L}_{0}^{(q)}
&=& -\sqrt{[4]/[2]}\left[T^\dagger\times\widetilde{T}\right]_{1 0}^{(1/q)}
\nonumber\\
&=& -T^\dagger_{+1}\widetilde{T}_{-1}
+ (q-q^{-1}) T^\dagger_{0}\widetilde{T}_{0}
+ T^\dagger_{-1}\widetilde{T}_{0}
\end{eqnarray}
	we obtain
\begin{equation}\label{eq:ha03}
X_{0}^{(q)} + q {\cal L}_{0}^{(q)}
= -[2] T^\dagger_{+1}\widetilde{T}_{-1}
+ q^2 T^\dagger_{0}\widetilde{T}_{0}
\end{equation}
	Since
\begin{equation}\label{eq:ha04}
T^\dagger_{+1}\widetilde{T}_{-1} = \frac{[2N_{+}]}{[2]} q^{-2N_{+}+2N+1} ,
\qquad
T^\dagger_{0}\widetilde{T}_{0} = [N_{0}] q^{-4N_{+}-N_{0}+2N-1}
\end{equation}
	we get upon summation
\begin{equation}\label{eq:ha04a}
X_{0}^{(q)} = -q {\cal L}_{1 0}^{(q)} + [N+L_{0}] q^{N-L_{0}+1}
\end{equation}
	and, after some calculations,
\begin{equation}\label{eq:ha05}
{}^qH_{3dim}= \hbar \omega_{0} X_{0}^{(q)} =
\hbar \omega_0\left\{ [N] q^{N+1}
- \frac{q(q-q^{-1})}{[2]} C_2^{(q)}\right\}
\end{equation}
	The eigenvalues of such a $q$-deformed Hamiltonian are
\begin{equation}\label{eq:ham04}
\fl {}^qE_{3dim} = \hbar\omega_0 \left\{[n]q^{n+1}-\frac{q(q-q^{-1})}{[2]}
[\ell][\ell+1] \right\}, \qquad \ell = n,n-2,\ldots,0~\mbox{or}~1
\end{equation}
	In the $q\to 1$ limit we have
	$\lim_{q\to1} {}^qE_{3dim}=\hbar \omega_0 n$,
	which coincides with the classical result.
	We shall note that, the expression (\ref{eq:ha05}) can also be
	represented in the form
\begin{equation}\label{eq:ham06}
{}^qE_{3dim}= \hbar \omega_{0} \frac{q^{n+1}}{[2]}
\left\{q^{\ell+1}[n-\ell] + q^{-\ell-1} [n +\ell]\right\}
\end{equation}
	but the advantage of the form (\ref{eq:ham04}) is that, the
	vibrational ($[n]$) and the rotational ($[\ell][\ell+1]$) degrees of
	freedom are clearly separated.

	For small values of the deformation parameter $\tau$, where
	$q=e^{\tau}$, we can expand (\ref{eq:ham04})
	in powers of $\tau$ obtaining
\begin{eqnarray}\label{eq:ham07}
{}^qE_{3dim}&=&\hbar\omega_0\; n
-\hbar\omega_0 \tau \left(\ell(\ell+1) - n(n+1)\right) \nonumber\\
&&-\hbar\omega_0 \tau^2 \left(\ell(\ell+1) - \frac{1}{3}n(n+1)(2n+1)\right)
+{\cal O}(\tau^3)
\end{eqnarray}
	To leading order in $\tau$ the expression (\ref{eq:ham07})
	closely resembles the one giving the energy eigenvalue
	of the MHO\footnote{
Indeed, these eigenvalues are
\[ E_{nl} = \hbar \omega n - \hbar \omega \mu^{\prime} \left(
l(l+1) - \frac{1}{2}n(n+3) \right) \]
where $\mu^{\prime}$ is allowed to vary form shell to shell.}
	(\ref{eq:i01}) (if one neglects the spin-orbit term).
	Quantitatively, one obtains a good fit of
	the energy spectrum produced by the MHO potential (\ref{eq:i01}) if
	we choose for $\hbar\omega_{0}$ and $\tau$ of the $q$-HO the values
	$\hbar \omega_{0} =0.94109$ and $\tau=0.021948$
	(we have assumed $\hbar\omega=1$ for MHO and
	{\it neglected} the spin-orbit term).
	The value of $\tau$ is close to those
	adopted for $\mu^{\prime}=k\mu$. Indeed the MHO fit
	in the $^{208}_{82}Pb$ region yields \cite{BR85,NR95}
	$\mu^{\prime}=0$ for $N=2$, $\mu^{\prime}= 0.0263$ for $N=3$,
	$\mu^{\prime} = 0.024$ for $N \geq 4$ and
	for $k$, fixed by the condition that the observed order of
	sub-shells be reproduced, $k=0.08$ for $N=2$,
	$k=0.075$ for $N=3$ and $k=0.07$ for $N \geq 4$.

	It appears surprising at first that $\hbar \omega_0$
	differs slightly from $1$. Indeed the correction term $n(n+1)$
	in the $q$-HO spectrum is slightly different from the
	corresponding piece $n(n+3)/2$ in the MHO. Some
	compression of the $q$-HO spectrum is therefore unavoidable.
	From a physical point of view, one may say that the
	mean radius of the deformed oscillator is slightly larger than
	the radius of the classical isotropic HO.

	MHO and $q$-deformed 3-dim HO spectra are compared in figure 1. It
	should be noted that the constant $\mu'$ in the modified harmonic
	oscillator potential takes different values for shells with different
	values of $N$.  In the $q$-deformed model the parameters $\tau$ and
	$\hbar\omega_0$ have  the same values for all the shells.  The
	comparison shows that the $q$-deformation of the 3-dimensional harmonic
	oscillator effectively reproduces the non-locality and the
	``deformations'' induced in the MHO-model through the terms
	${\bf L}^2$,~$\langle{\bf L}^2\rangle_N$
	and the variability of $\mu^{\prime}$ with the shell number $N$.

\section{\normalsize\bf
$q$-deformed 3-dimensional harmonic oscillator
with $q$-deformed spin-orbit term}

	For a full comparison with the MHO and a more realistic description
	of the single-particle spectrum, we need to include a $q$-deformed
	spin-orbital term ${\cal L}^{(q)}\cdot{\cal S}^{(1/q)}$ in the
	$q$-deformed harmonic oscillator Hamiltonian ${}^qH_{3dim}$. To this
	purpose we shall introduce the spin operators $S_{+},S_{0},S_{-}$,
	which are elements of another (independent) $su_q(2)$ algebra. These
	operators satisfy the commutation relation (\ref{eq:s07}), i.e.
\begin{equation} \label{eq:ls01}
[S_{0},S_{\pm}] = \pm S_{\pm} \qquad\qquad
[S_{+},S_{-}] = [2S_{0}]
\end{equation}
	and act in the {\it two-dimensional} representation space of
	this algebra. The orthonormalized basis vectors of this space will be
	denoted as always by $|\frac{1}{2} m_s\rangle_q$.

	We now define the $q$-deformed total angular momentum as \cite{Sm}
\begin{eqnarray} \label{eq:ls02a}
J_{0} &=& L_{0} + S_{0} \\
J_{\pm} &=& L_{\pm} q^{S_{0}} + S_{\pm} q^{-L_{0}}  \label{eq:ls02b}
\end{eqnarray}
	where the $q$-deformed orbital angular momentum is given by
	(\ref{eq:s06}).
	The operators (\ref{eq:ls02a}-\ref{eq:ls02b}) satisfy the same
	commutation relations as the operators of the $q$-deformed orbital
	angular momentum (\ref{eq:s07}) and spin (\ref{eq:ls01}), and the
	corresponding expression for the Casimir operator of the algebra
	(\ref{eq:ls02a}-\ref{eq:ls02b}) can be written in the form
\begin{equation} \label{eq:ls02c}
C_{2,J}^{(q)} = J_{-}J_{+} + [J_0][J_0+1]
\end{equation}
	As in the case of the $q$-deformed orbital momentum we shall consider
	$^qJ^2\equiv C_{2,J}^{(q)}$ as the total angular momentum ``square''.

	The common eigenvectors of $C_{2,J}^{(q)}, C_{2,L}^{(q)}$
	and $C_{2,S}^{(q)}$ can be written in the usual form
\begin{equation} \label{eq:ls03}
|n (\ell \case{1}{2}) j m\rangle_q = \sum_{m_{\ell},m_{s}}
{}^qC_{\ell m_\ell, \case{1}{2} m_s}^{j m}
|n \ell m_\ell\rangle_q |\case{1}{2} m_s\rangle_q
\end{equation}
	where ${}^qC_{\ell m_\ell, s m_s}^{j m}$ are the
	$q$-deformed Clebsch-Gordan coefficients.

	We define the spin-orbital term as a scalar product, according
	to the definition (\ref{eq:scalar})
\begin{eqnarray} \label{eq:ls04}
{\cal L}^{(q)}\cdot {\cal S}^{(1/q)}
&=& \sum_{M=0,\pm1} (-q)^{M} {\cal L}_{M}^{(q)}{\cal S}_{-M}^{(1/q)} =
\nonumber\\
&=& \frac{1}{[2]}\left\{ C_{2,J}^{(q)} - C_{2,L}^{(q)} - C_{2,S}^{(q)}
- \frac{(q-q^{-1})^2}{[2]}C_{2,L}^{(q)}C_{2,S}^{(q)} \right\}
\end{eqnarray}
	where ${\cal S}^{(1/q)}$ is a vector operator according to the
	algebra \eref{eq:ls01} and it is constructed from the
	$q$-deformed spin operators taking into account the rule
	\eref{eq:Ltensor}, but for deformation parameter $1/q$.
	The Hamiltonian, which we suggest for the $q$-deformed
	3-dim harmonic oscillator with spin-orbit interaction is
\begin{eqnarray} \label{eq:ls05}
\fl && {}^qH = \hbar\omega_{0}
\left\{ X_0^{(q)} - \kappa [2]{\cal L}^{(q)}\cdot {\cal S}^{(1/q)}\right\} =\\
\fl &&= \hbar\omega_{0}\left\{
[N] q^{N+1} - \frac{q(q-q^{-1})}{[2]} C_{2,L}^{(q)}
-\kappa\left( C_{2,J}^{(q)} - C_{2,L}^{(q)} - C_{2,S}^{(q)}
- \frac{(q-q^{-1})^2}{[2]}C_{2,L}^{(q)}C_{2,S}^{(q)} \right)\right\}
\nonumber
\end{eqnarray}
	In \eref{eq:ls05} the factors have been chosen in accordance with
	the usual convention in classical (spherically symmetric) shell model
	with spin-orbit coupling.
	The eigenvectors of this Hamiltonian are given by (\ref{eq:ls03})
	and the corresponding eigenvalues are
\begin{eqnarray} \label{eq:ls06}
\fl  && {}^qE_{n \ell j} = \hbar\omega_{0}\left\{
[n] q^{n+1} - \frac{q(q-q^{-1})}{[2]} [\ell][\ell+1] \right. \nonumber\\
&&  \left.   -\kappa\left( [j][j+1] - [\ell][\ell+1] -
{\textstyle [\frac{1}{2}][\frac{3}{2}]}
- \frac{(q-q^{-1})^2}{[2]} [\ell][\ell+1]
{\textstyle [\frac{1}{2}][\frac{3}{2}]} \right)\right\}
\end{eqnarray}

	In order to compare the expression for $q$-deformed $L-S$ interaction
	with the classical results we give the expansion of the
	expectation value (\ref{eq:ls04}) in series of $\tau$,~
	$(q=e^\tau, \tau\in{\bf R})$
\begin{eqnarray}\label{eq:ls07}
\fl && {}_q\langle n(\ell \case{1}{2})jm|{\cal L}^{(q)}\cdot {\cal S}^{(1/q)}
|n(\ell \case{1}{2})jm \rangle_q = \\
&& = \cases{ \frac{1}{2} \ell
\left(1+\frac{\tau^2}{6}(4 \ell^2- 7) + \ldots \right) &
if $j=\ell+1/2$ \\                  \nonumber
 -\frac{1}{2}(\ell+1)\left(1+\frac{\tau^2}{6}(4 \ell^2 + 8\ell -3)
+ \ldots \right) & if $j=\ell-1/2$}
\end{eqnarray}
	It can be easilly seen that expression (\ref{eq:ls04}) introduces
	some corrections to the classical expressions for the spin-orbit
	interaction, which are proportional to $\tau^2$. These corrections
	are small for the light nuclei, where the values of $\ell$ are small,
	but they are not negligable for the heaver nuclei,
	where the shells with bigger values of $\ell$ are important.

	On the left side of figure 2 are shown the levels of MHO
	with spin-orbit term, considered again
	in the $^{208}_{82}Pb$ region \cite{NR95} and the constants in the
	expression \eref{eq:i01} $\mu'=k\mu$ and $k$ are the same as
	in figure 1. On the right side of the
	same figure are shown the levels of the $q$-deformed 3-dimensional
	harmonic oscillator with $q$-deformed spin-orbit interaction.
	It should be noted that the levels of the
	$q$-deformed oscillator are calculated with the same values of the
	deformation parameter $q$ for the oscillator term and the $q$-deformed
	$L-S$ term, and, as well as, with the same constant $\kappa$ of the
	spin-orbit interactiion for all shells.

\section{\normalsize\bf
Conclusions}

       The results of this paper demonstrate, that $q$-deformed algebras
       provide a natural background for the description of "deformable"
       physical systems.

       In particular, we have shown, that in the enveloping algebra of
       $so_q(3)$ there exists an $so_q(3)$ scalar operator, built of
       irreducible vector operators according to the reduction
       $u_q(3)\supset so_q(3)$, for the case of the most symmetric
       representations of $u_q(3)$. In the limit $q\rightarrow 1$ this
       operator coincides with the Hamiltonian of the 3-dimensional harmonic
       oscillator, while, in the case of small deformations of the algebra,
       its spectrum reproduces the spectrum of the 3-dimensional modified
       harmonic oscillator (with, or without a spin-orbit interaction),
       making transperent the connection of our model with shell model
       calculations.

       It should be emphasized, that in our case the parameters of the
       $q$-deformed model are the same for all shells, while in the modified
       harmonic oscillator the parameters vary from shell to shell.

       The quantitative comparison of both models shows, that the
       $q$-deformation of the 3-dimensional harmonic oscillator effectively
       reproduces the "non--locality" and "deformations", introduced in the
       model of the modified harmonic oscillator by means of additional
       correction terms.

       The application of the $q$-deformed 3-dimensional harmonic oscillator
       model to real nuclei will be considered in another paper, which is now
       in progress.

\ack
       The authors (PR) and (NL) acknowledge the financial support of the
       Italian Ministry of University Research and Technology (MURST) and of
       the Istituto Nazionale di Fisica Nucleare (Italy).
       The authors (PR) and (RR) has been supported by the Bulgarian Ministry
       of Science and Education under contracts $\Phi$-415 and $\Phi$-547.

\end{document}